\documentclass[11pt,a4paper]{article}
\usepackage{jcappub}

\title{On the emergence of the Lorentz signature in an expanding universe}
\author[a,b]{Angelo Tartaglia}

\affiliation[a]{Politecnico di Torino,\\
corso Duca degli Abruzzi 24, 10129 Torino, Italy}
\affiliation[b]{INFN, Sezione di Torino, \\
via Pietro Giuria 1, 10126 Torino, Italy}

\emailAdd{angelo.tartaglia@polito.it}

\abstract{A mechanism producing the transition from an Euclidean to a Minkowski
manifold is described. A global Robertson-Walker symmetry is assumed from
the large scale data of the visible universe. Allowing for the strain of the
manifold as an additional field in the Lagrangian, we interpret the symmetry
as a consequence of a global texture defect. The additional term gives rise
to a boundary dividing the manifold into an Euclidean plus a Lorentzian
region. It is also shown that the presence in the early epoch of homogeneous matter/energy fields
preserves the horizon and the signature change across it. The horizon has properties
much similar to the ones of the Big Bang of the Standard Model, including
the need for a phase transition of the scalar field producing particles and
fields as we know them now.}


\begin{document}

\maketitle

\section{Introduction}

The commonly assumed background of special relativity is a Minkowski
space-time, i.e. a flat four-dimensional manifold equipped with a Lorentzian
signature. When generalizing the theory of relativity (GR) by the
introduction of curvature in the actual space-time, the Minkowski manifold
is typical of all tangent spaces to the curved manifold. Minkowski
space-time as such does in a sense not exist, but is the asymptotic form of
any real space-time when all kinds of matter/energy are taken out. This
simple view, however, hides a puzzling feature. A Minkowski manifold is not
the most general undifferentiated flat four-dimensional manifold, because of
the light cones, i.e. of the Lorentzian signature. The structure of the
light cones picks out a bunch of directions stemming from any given event in
the manifold (the time-like worldlines) which cannot be confused with the
rest. Where does this symmetry diminution come from? Indeed the most general
four-dimensional manifold should be Euclidean: perfect isotropy and
homogeneity. This is the question for which I shall try to find an answer in the present work.

The problem of the signature of space-time has already been considered in the past from different viewpoints and with different motivations. A role of Euclidean signature geometries in four dimensions has been introduced, for instance, in quantum cosmology where Euclidean path integrals over geometries have been generalized from flat quantum field theory to gravity and the universe \cite{hartle} \cite{hawking}. In a quantum approach the transition from Euclidean to Lorentzian signature is achieved by analytic continuation in the complex domain or formally just by means of a Wick's rotation (introduction of imaginary time $t\rightarrow it$). The existence of a Euclidean patch with the geometry of a four-dimensional half sphere was intended, in Hawking's approach, to remove the singularity in the remote past of the universe. How a real classical interpretation of the results emerges out of the quantum methods is however a non-trivial matter.

The approach I will adopt in this paper is entirely classical \emph{ab initio}. Even from a classical viewpoint the possibility of having different signatures in different regions of space-time has been discussed in the literature, especially in the 90's of the past century \cite{ellis1}\cite{ellis2}. The discussion was soon concentrated on the general constraints posed by a possible signature flip somewhere in the manifold and on different viable approaches \cite{embacher}\cite{hellaby}. The continuity conditions at the border between the Euclidean and the Lorentzian domains were analyzed together with the conservation laws on the boundary \cite{dray}\cite{sumeruk}. Though establishing the compatibility conditions for the co-existence of Euclidean and Lorentzian domains, the actual presence of Euclidean regions is then left to further theories, without being considered as a logical implication of the symmetry breaking manifested by the presence of the light cones. The purpose of the present paper is precisely to provide a consistent classical framework wherein the matching between Euclidean and Lorentzian signature is obtained "naturally".

The question is: can we envisage a mechanism by which, pouring matter/energy (or something
else) in a Euclidean manifold, the Lorentzian signature appears? In fact a way to induce a peculiar
symmetry (to reduce the total symmetry) in a continuum is to introduce a
defect in the sense used when describing material continua\cite{volterra}%
\cite{punti}. This is what I meant by "something else" in the line above.
However it is not clear how a defect can convert a ($++++$) into a ($+---$)
signature. In order to find the way out I shall follow a path outlined in
\cite{cqg} and verified by various cosmological tests \cite{test}; for it I
shall use the name Strained State Cosmology (SSC) or Strained State Theory (SST). It consists in assuming
that space-time behaves as a continuous deformable medium in four
dimensions, extending the usual general relativistic approach by the
introduction in the Lagrangian density of empty space-time of an "elastic
potential" contribution built from the strain tensor; the final Lagrangian
looks very similar to the ones used in the so called "massive gravity" but
does indeed not coincide with them. The strain tensor of the manifold is
intended as being half the difference between the actual metric tensor and
the metric tensor of a reference undeformed manifold. According to the
considerations I made above the reference manifold should be an Euclidean
one. Mentioning two metric tensors gives the impression that I am presenting a bimetric theory. This however is not the case. Only one of the manifolds is actually existing: the one corresponding to our space-time (the natural manifold). The reference manifold is not present anywhere; it is not even a background. It simply is part of a logical description in which the universe is thought to behave as a deformed continuum. For this reason the mentioned metric tensor of the Euclidean manifold is no metric at all in the natural manifold.

The piece of evidence from which we start is that we know (or at least we
think we know) that the universe, at scales of hundreds of Mpc or higher, is
described by a Friedman-Lema\^{\i}tre-Robertson-Walker model, i.e. it has a
Robertson-Walker (RW) symmetry (isotropic expansion of a homogeneous space).
Since the presence of matter \textit{per se} does not motivate the RW
symmetry I shall attribute the symmetry fixing to the presence of a global
defect, somewhere in the manifold.

In practice the initial assumptions will be:

\begin{itemize}
\item an action integral of the empty space-time like the following:%
\begin{equation}
\int \left[ R+\frac{1}{2}\lambda \varepsilon ^{2}+\mu \varepsilon _{\alpha
\beta }\varepsilon ^{\alpha \beta }\right] \sqrt{-g}d^{4}x  \label{azione}
\end{equation}%
where $\lambda $ and $\mu $ are the Lam\'{e} coefficients of space-time
(fixed parameters) and $\varepsilon _{\mu \nu }=\frac{1}{2}\left( g_{\mu \nu
}-E_{\mu \nu }\right) $ is the strain tensor determined with respect to the
symmetric tensor $E_{\mu \nu }$ that would correspond to the Euclidean metric on the reference manifold; it is also $\varepsilon =\varepsilon
_{\alpha }^{\alpha }$ where all the raising and lowering of indices is performed by means of the unique metric tensor $g_{\mu\nu}$; no assumption is made about $\varepsilon _{\mu \nu }$
being small or not with respect to $E_{\mu \nu }$; $R$ plays the role of
dynamical term for the strain tensor components;

\item a defect inducing a global RW symmetry.
\end{itemize}

Under these conditions we shall see that a Euclidean signature in a domain of the natural manifold
matches a Lorentzian one in correspondence with an horizon in the manifold. We
shall also see that the presence of matter/energy does not spoil the effect
I have just mentioned, provided the additional ingredient, under the horizon,
is in the form of a completely homogeneous field. Of course such a field
should then undergo a phase transition giving rise to the ingredients of the
universe we observe now.

The SST whose essence has been outlined above is indeed a metric theory on a Riemannian manifold which admits everywhere, excluding singularities, a flat tangent space-time. In the Lorentzian domain the tangent space is Minkowskian; in practice this tells us that the theory preserves the principle of equivalence and recovers locally the special relativity.

\section{The expanding space-time}

The assumptions presented in the introduction imply that the line element
for the actual space-time (natural line element) has the form:

\begin{equation}
ds^{2}=d\tau ^{2}-a^{2}\left( \tau \right) \left( dx^{2}+dy^{2}+dz^{2}\right)
\label{naturale}
\end{equation}

The meaning of the symbols is the traditional one in GR and Cartesian
coordinates have been chosen for simplicity. The time dependence of the
scale factor $a$ implies the global curvature of the manifold and indeed the
RW global symmetry. The space has been assumed to be flat, since this is
what we conclude at the moment from the observation of the CMB. In any case
a space with positive or negative curvature would not modify the
considerations I am about to make and the final conclusions.

The line element on the reference Euclidean manifold is

\begin{equation}
ds_{r}^{2}=b^{2}\left( \tau \right) d\tau ^{2}+dx^{2}+dy^{2}+dz^{2}
\label{riferim}
\end{equation}%
where a gauge freedom has been allowed in the choice of the "time" coordinate
identifying the correspondence between points on the two manifolds; of course this coordinate is just as space-like as the others in the flat Euclidean manifold. This
gauge (in practice the lapse function $b$) is motivated by the fact that the only variable entering the curvature
of the RW manifold is $\tau$. In the discussion by other authors of the change of signature problem the analog of the gauge function $b$ with the corresponding degree of freedom is introduced in the form of a lapse function in the natural line element (the only one they consider)\cite{ellis2}.

From the definition of the natural and the reference line elements we immediately obtain the strain tensor:

\begin{equation}
\begin{array}{c}
\varepsilon _{00}=\frac{1-b^{2}}{2}\smallskip \\
\varepsilon _{xx}=-\frac{a^{2}+1}{2}\smallskip \\
\varepsilon _{yy}=-\frac{a^{2}+1}{2}\smallskip \\
\varepsilon _{zz}=-\frac{a^{2}+1}{2}\smallskip%
\end{array}
\label{strain}
\end{equation}

The second order scalars associated with the strain tensor are:

\begin{equation}
\varepsilon ^{2}=\left( \frac{1-b^{2}}{2}+3\frac{a^{2}+1}{2a^{2}}\right) ^{2}
\label{traccia}
\end{equation}

and

\begin{equation}
\varepsilon _{\alpha \beta }\varepsilon ^{\alpha \beta }=\left( \frac{1-b^{2}%
}{2}\right) ^{2}+3\left( \frac{a^{2}+1}{2a^{2}}\right) ^{2}  \label{doppio}
\end{equation}

Introducing (\ref{traccia}) and (\ref{doppio}) into (\ref{azione}) we obtain
the Lagrangian density

\begin{eqnarray*}
\mathcal{L} &\mathcal{=}&\left[ R+\frac{1}{2}\lambda \varepsilon ^{2}+\mu
\varepsilon _{\alpha \beta }\varepsilon ^{\alpha \beta }\right] \sqrt{-g} \\
&=&a^{3}\left( \frac{6}{a^{2}}\dot{a}^{2}+\frac{1}{4}\left( \frac{\lambda }{2%
}+\mu \right) \left( 1-b^{2}\right) ^{2}+\frac{3}{4}\left( \frac{\lambda }{2}%
3+\mu \right) \left( \frac{a^{2}+1}{a^{2}}\right) ^{2}+\frac{3}{4}\lambda
\left( 1-b^{2}\right) \frac{a^{2}+1}{a^{2}}\right)
\end{eqnarray*}%

An integration by parts has been made in order to reduce the second order derivative
present in the scalar curvature. The $b$ function appears only in the terms containing the strain; the
Euler-Lagrange equation for it reduces to $\partial \mathcal{L}/\partial b=0$%
. Explicitly we find:

\begin{equation}
\lambda \left( \frac{1-b^{2}}{2}+\frac{3}{2}\frac{a^{2}+1}{a^{2}}\right)
+2\mu \left( \frac{1-b^{2}}{2}\right) =0  \label{nondin}
\end{equation}%
whose solution is:

\begin{equation}
b^{2}=2\frac{2\lambda +\mu }{\lambda +2\mu }+\frac{3}{a^{2}}\frac{\lambda }{%
\lambda +2\mu }  \label{bsol}
\end{equation}%
$\allowbreak $

For consistency ($b^{2}>0$ for any $a$) this solution requires that it be:%
\begin{eqnarray*}
\mu &>&-\frac{\lambda }{2}\text{ if }\lambda >0 \\
\mu &<&-\frac{\lambda }{2}\text{ if }\lambda <0
\end{eqnarray*}

The dynamics is fixed by the Euler-Lagrange equation for $a$, after
introducing (\ref{bsol}) \cite{test}:%
\begin{equation}
2\left( 2\ddot{a}a+\dot{a}^{2}\right) -\frac{\mu }{2a^{2}}\frac{2\lambda
+\mu }{\lambda +2\mu }\left( 3a^{4}+2a^{2}-1\right) =0  \label{ELeqE}
\end{equation}

A first integral of (\ref{ELeqE}) is obtained by the energy condition:

\[
W=6a\dot{a}^{2}-\frac{3}{2a}\mu \frac{\left( a^{2}+1\right) ^{2}}{\lambda
+2\mu }\left( 2\lambda +\mu \right)
\]%
from which the Hubble parameter can be written

\[
H^{2}=\frac{\dot{a}^{2}}{a^{2}}=\frac{1}{6a^{3}}\left( W+\frac{3}{2}\mu
\frac{2\lambda +\mu }{\lambda +2\mu }\frac{\left( a^{2}+1\right) ^{2}}{a}%
\right)
\]

In order to reproduce GR when $\mu =0$, space-time is empty and space is
flat it must be $W=0$ so that

\begin{equation}
\frac{\dot{a}}{a}=\pm \sqrt{\frac{\mu }{4}\frac{2\lambda +\mu }{\lambda
+2\mu }}\frac{\left( a^{2}+1\right) }{a^{2}}  \label{hub2}
\end{equation}

In the absence of matter and excluding special values for the Lam\'{e}
coefficients one has continued expansion or contraction only. Let us choose
expansion, then let us solve (\ref{hub2}). It comes:

\begin{equation}
a=\sqrt{Ce^{\sqrt{\mu \frac{2\lambda +\mu }{\lambda +2\mu }}\tau }-1}
\label{sola}
\end{equation}%
where only positive values of $a\allowbreak $ have been considered and $C$
is an integration constant. The variable $\tau $ is the distance from the
origin thought as being the seat of a "defect" inducing the global RW
symmetry; it is measured along a geodesic line everywhere perpendicular to the equal strain hypersurfaces; in the Lorentzian domain $\tau $ corresponds to the cosmic time.
$C$ can be interpreted as being a feature, or "strength", of the
defect; the situation for $C=0$ corresponds to the flat reference Euclidean
manifold.

For $\tau =0$ it is:%
\[
a\left( 0\right) =\sqrt{C-1}
\]

A global Lorentzian signature in late times would imply $C>0$, however if $%
0<C<1$ we find an initial era during which the signature of the natural
manifold is Euclidean. Then at

\[
\tau _{h}=\sqrt{\frac{\lambda +2\mu }{\mu \left( 2\lambda +\mu \right) }}\ln
\frac{1}{C}
\]%
we find a horizon where $a\left( \tau _{h}\right) =0$ and out of which (for $%
\tau >\tau _{h}$) a Lorentzian signature emerges.

\section{The effect of matter}

Let us now allow "matter" to be present. In the Euclidean era a consistent
presence could be in the form of a uniformly distributed field; for simplicity let us assume that its density scales as $a^{4}$, as afterwards would be the case for
radiation. Of course if we want to have ordinary matter at late time we need
some mechanism giving rise, on our side of the horizon, to matter terms
scaling as $a^{3}$; let us say that some "phase transition" is needed or that a matter component should be present already in the Euclidean era. In any case, as we shall see, the relevant domain is close to the horizon on both sides and there the dominant term is the $\sim 1/a^4$.

In order to study the behaviour before and in the vicinity of the horizon let
us introduce in (\ref{hub2}) an additional term that will lead to%
\begin{eqnarray}
\frac{\dot{a}}{a} &=&\sqrt{B^{2}\frac{\left( a^{2}+1\right) ^{2}}{a^{4}}+%
\frac{K^{2}}{a^{4}}}  \label{equaz} \\
B^{2} &=&\frac{\mu }{4}\frac{2\lambda +\mu }{\lambda +2\mu }
\end{eqnarray}

From (\ref{equaz}) we have

\[
\int \frac{d\left( a^{2}\right) }{2\sqrt{B^{2}\left( a^{2}+1\right)
^{2}+K^{2}}}=\tau +T
\]%
i.e.

\[
\frac{1}{2B}\sinh^{-1} \left( \frac{B\left( a^{2}+1\right) }{K}\right) =\tau
+T
\]%
$T$ is an integration constant that can be equalled to zero without loss of
generality. Finally:

\begin{equation}
a^{2}=\frac{K}{B}\sinh 2B\tau -1  \label{solk}
\end{equation}

We find a Euclidean era ending at

\begin{equation}
\tau _{h}=\frac{1}{2B}\sinh^{-1}\left( \frac{B}{K}\right)  \label{hk}
\end{equation}%

The variable $\tau_{h}$, under the boundary between the Euclidean and the Lorentzian domains, is not an observable. Our cosmic time is measured starting from the horizon, i.e. it is relative to $\tau_{h}$. Afterwards
the usual expansion in a Lorentzian space-time follows. If the content
of the universe, after $\tau =\tau _{h}$, would remain limited to the $%
K/a^{4}$ component, the expansion would be continuous and accelerated. The
situation changes if a component appears, proportional to $1/a^{3}$. Should that component have been present right from the beginning in Eq. \ref{equaz} I do not expect qualitative changes in the final result, but simply a shift in the position of the horizon.

If we introduce (\ref{sola}) into (\ref{equaz}) we see that the latter is
satisfied provided we let $K\rightarrow 0$ thus recovering the solution for
the empty space-time with a defect inducing the RW symmetry. Of course for $%
K\rightarrow 0$ also a pure Euclidean manifold is a solution.

\section{Junction conditions at the horizon}

Let us have a closer look to what happens at the transition. Both in the case of an empty space-time and in presence of matter/energy the metric tensor is singular at the horizon since the scale factor $a$ vanishes there. However $a^2$ passes smoothly from negative to positive values while crossing the horizon; there are no jumps or discontinuities at any order. Of course the metric inverse is undefined on the horizon and this fact produces the divergence and a discontinuity in some of the Christoffels. The relevant quantity is the Einstein tensor $G_{\mu\nu}$. Its elements diverge when $a=0$ but of course the four-divergence is identically zero: $G_{\mu;\nu}^{\nu}=0$. The element $G_{00}$, whose expression, in presence of a matter content, is
 \[
G_{00}=3B^{2}K^{2}\frac{cosh^{2}{2B\tau}}{(ksinh{2B\tau}/B-1)^{2}},
\]%
has the same limit on the left and on the right of the junction. This is one of the conditions to be satisfied, according to the discussion that can be found in \cite{dray}, in order not to violate the usual conservation of matter. By the way, this result is typical of Friedman-Lema\^{\i}tre-Robertson-Walker cosmologies with zero space curvature ($k=0$), as it is the case for the model I am discussing here.

\section{Conclusion}

As we have seen, it is possible to think of a four-dimensional manifold
where both the Euclidean and the Lorentzian signature are present in
different regions. The boundary between the two regions is smoothly crossed
if the global manifold has a Robertson-Walker symmetry and the given
symmetry can be there as a consequence of a texture defect. The simplest
representation of the configuration of the manifold is obtained when the
$\tau $ variable is chosen as originating from the defect and increasing along
world-lines perpendicular to the equal strain hypersurfaces of the manifold.
The shape of the latter hypersurfaces depends on the geometry of the defect;
if the space in our universe is flat, as apparently it is, so has to be the
singular submanifold corresponding to the defect. The boundary between the
Euclidean and the Lorentzian areas is the hypersurface corresponding to a
null scale factor, $a=0$; this boundary is called a "horizon" from which the
Lorentzian signature emerges, in analogy with the Schwarzschild horizon for
a cylindrical stationary space-time (spherical in space).

Describing the situation I have now and then used terms as \emph{evolution}, \emph{transition}, even \emph{time}. Of course this way of depicting the manifold is, strictly speaking, inappropriate in the Euclidean domain. There, in fact, there is no time; all dimensions are space-like; nothing \emph{propagates}. In the Euclidian domain the $\tau $ variable is an affine parameter along incomplete geodesics bounded by the cosmic defect on one side. The geodesics, besides being incomplete, are chosen so that they are everywhere perpendicular to the equal strain hypersurfaces of the manifold. It is only on the Lorentzian side of the boundary between the two domains that $\tau $ acquires the familiar role of \emph{cosmic time}. By the way in a consistent fully geometric view any description in terms of \emph{evolution} is somehow inappropriate: there is just one fourdimensional manifold described in terms of Gaussian coordinates, curved and locally warped according to a global symmetry and the local distribution of matter/energy. The manifold has two regions of different signature, matching each other in a reasonably smooth way on a three-dimensional boundary. The term \emph{evolution} is what we use for the way we label $3+1$ foliations of the manifold in the Lorentzian domain. Even though we would not use \emph{evolution} for it, a similar labeling for an analogous foliation is possible also on the Euclidean side: it would not be constrained by the light cones, but would be suggested by the symmetry. Of course this is so in an entirely classical approach, but so far the puzzle of the role of time and of some background in the attempts to quantize gravity remains unsolved.

It is important to remark that, even though the action integral (\ref{azione}%
) looks very much similar to the action for the so called "massive gravity",
it is however different. In fact the original approach due to Firtz and
Pauli \cite{FP} viewed (\ref{azione}) as a first order approximation of a
perturbative treatment, whereas in our case (\ref{azione}) is "exact".
Further developments of the massive gravity theory, introduced in order to
cure various inconveniences present in the original version, do consider
also non-linear approaches where the perturbative treatment is extended, in
principle, to all orders, however in practice they are bimetric theories,
where to the dynamical metric tensor a background non-dynamical metric is
added, and the latter is used for building many scalars of the theory \cite%
{hinter}. In the SSC instead, there exists just one metric and $E_{\mu \nu }$
is not a background and is not used as a metric tensor at all; raising and lowering of indices, then the construction of all scalars, are performed by means of the unique $g_{\mu\nu}$, just as in classical GR. Furthermore
the problems whose presence is still a matter of debate in the massive
gravity theories are absent from SSC, at least as far as the cosmic scale is
concerned. An immediate example is the absence, in the cosmological solution, of the so called vDVZ (van Dam-Veltman-Zakharov \cite{vDV}\cite{zakh}) discontinuity: when letting $\lambda$ and $\mu$ go to zero ($B$ go to zero) in (\ref{hub2}) and (\ref{equaz}) the shear GR cosmological solutions are obtained.

The features described so far pertain to the only manifold, without calling
in matter fields of any sort: in this case on the Lorentzian side we have an empty
space-time whose space expands for ever according to formula (\ref{sola}).
We have then considered the presence, in the Euclidean era, of matter/energy
in the form of a fully homogeneous radiation-like field (in four dimensions).
Provided the density of the field is small enough, the conclusion concerning
the presence of a horizon does not substantially change, as we see in
formula (\ref{hk}). However if we want to recover the present universe we
must allow for a phase transition occurring in the primordial field in
correspondence of the horizon, so that from it not only the Lorentzian
signature appears but also the abundance of fields and particles we
experience today. It is evident that a strong analogy exists between our
horizon and the Big Bang of the Standard Model. In our case, under the
horizon ("before" the Big Bang) we find an Euclidean era. Afterwards the
present approach based on the physical role of the strain of the manifold
has been successfully tested on a number of standard cosmological tests \cite%
{test}.

\end{document}